\documentstyle[aps,multicol,epsf]{revtex}

\begin{document}
\draft

\title{Grain Segregation Mechanism in Aeolian Sand Ripples}

\author{Hern\'an A. Makse}
\address{Schlumberger-Doll Research, Old Quarry Road, Ridgefield, CT
06877}

\date{Eur. Phys. J.-E, 1 January 2000} 
\maketitle
\begin{abstract}
Many sedimentary rocks are formed by migration of sand ripples.  Thin
layers of coarse and fine sand are present in these rocks,
and understanding how layers in sandstone are created has been a
longstanding question.
Here, we propose a mechanism for the origin of the most common layered
sedimentary structures such as inverse graded climbing ripple
lamination and cross-stratification patterns.  The mechanism involves
a competition between three segregation processes: {\it (i)}
size-segregation and {\it (ii)} shape-segregation during transport and
rolling, and {\it (iii)} size segregation due to different hopping
lengths of the small and large grains.
We develop a discrete model of grain dynamics
which incorporates the coupling between moving grains and the static
sand surface, as well as the different properties of grains, such as
size and roughness, in order to test the plausibility of this physical
mechanism.

\end{abstract}

\pacs{PACS: 81.05.Rm, 91.65.Ti}


\begin{multicols}{2}
\narrowtext

\section{Introduction}

\begin{figure}
\centerline{ \hbox{ \epsfxsize=8.cm \epsfbox{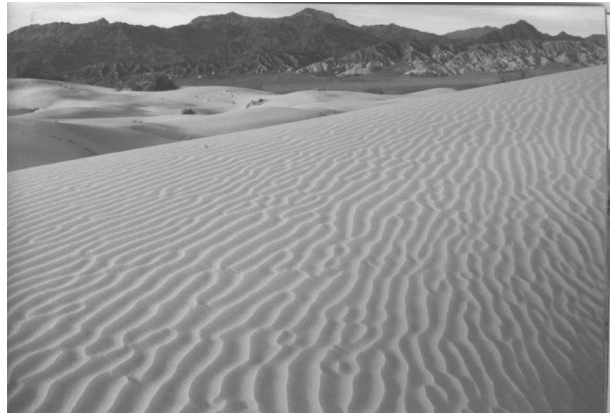} } }
\vspace{.5cm}
\caption{Wind sand ripples on a sandy surface in Death Valley National
Park, California.}
\label{death-valley}
\end{figure}

Subject to the effects of wind, a flat sandy surface is unstable and
evolves into a regular periodic pattern of wavelength of the order of
10 cm and height of a few centimeters (see Fig. \ref{death-valley})
\cite{bagnold,sharp,hunter1,origin,reineck,hunter2,allen,kocurek,schenk}.
A slight sand accumulation on the surface tends to expose grains on
the upwind stoss-side to the action of the wind, and shelter the
grains on the downwind lee-side. Dislodged grains at the stoss tend to
move toward the lee of the sand accumulation, the initial perturbation
is amplified, and small ripples develop, which migrate in the
direction of the wind.  Smaller ripples travel faster than larger
ripples--- since the migration velocity of ripples depends on the
amount of grains being transported during migration--- so that small
ripples merge with the larger ones.  Due to ripple merging, the
wavelength of the ripples grows in time as observed in the field
\cite{sharp,werner} and in wind tunnel experiments
\cite{seppala,anderson2}.

Due to the action of the wind, grains fly above the bed and strike the
ground at small angles and with velocities given by the wind velocity.
Successive impacts of grains are called {\it saltation}.  As a result
of bombardment of saltating grains, a number of ejected grains are
generated from the static bed which subsequently move by surface
creep.  These grains (called {\it reptating} grains) move distances
smaller than the typical saltating jump (which is controlled by the
wind velocity) until they are captured and become part of the bed.

\begin{figure}
\centerline{ \hbox{ \epsfxsize=8.cm \epsfbox{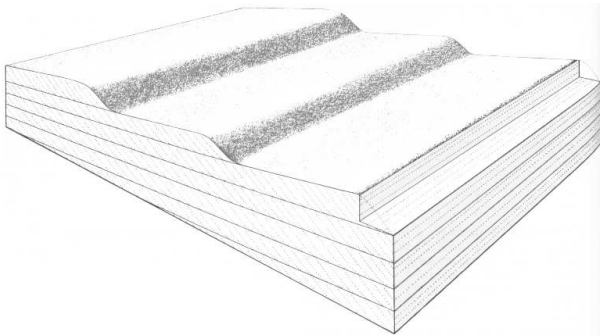} } }
\vspace{.5cm}
\caption{Diagram showing the climbing of the ripples and the formation
of cross-stratification patterns (from \protect\cite{reineck}).}
\label{diagram}
\end{figure}
\begin{figure}
\centerline{ \hbox{ \epsfxsize=9.cm \epsfbox{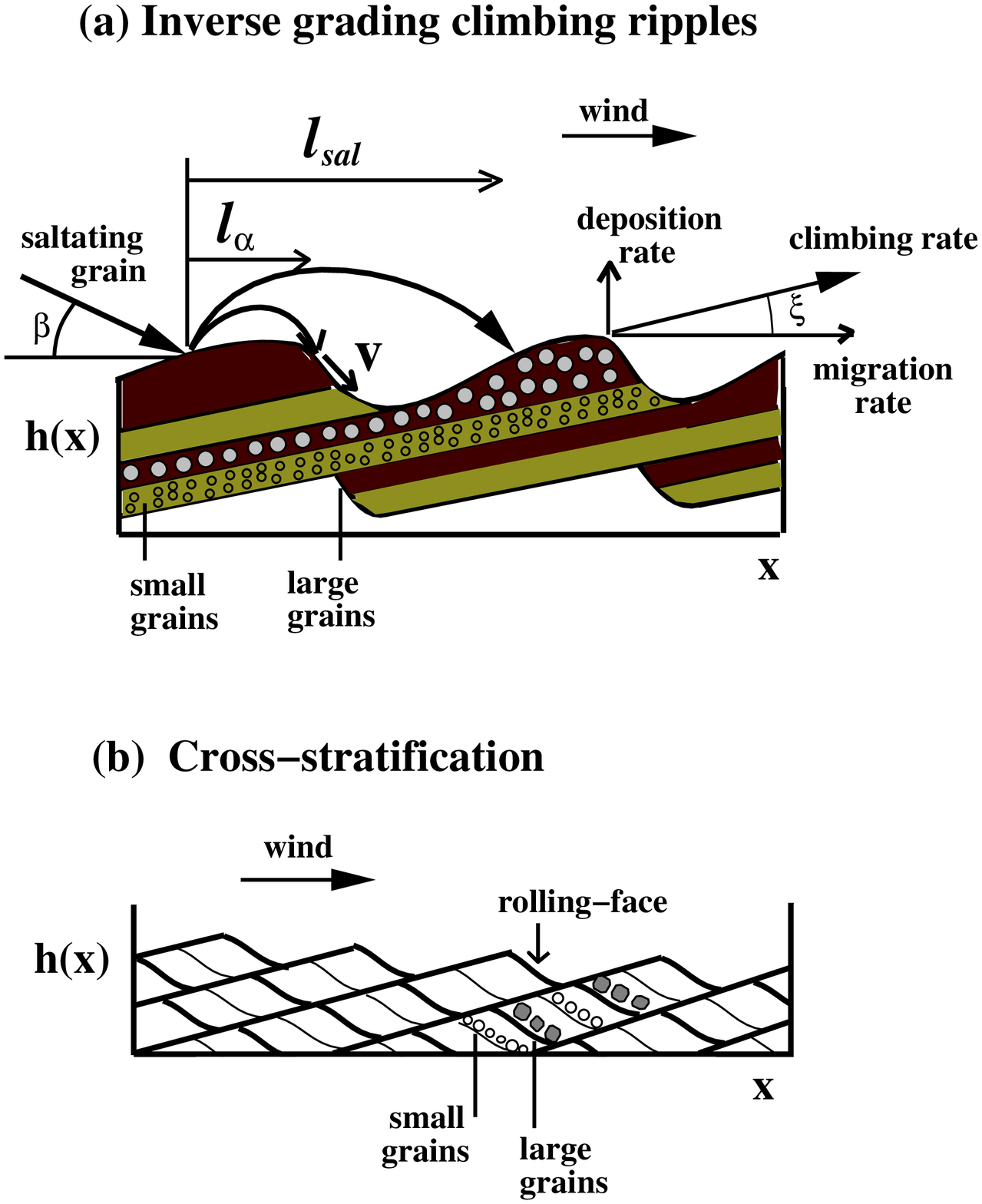} } }
\vspace{.5cm}
\caption{({\bf a}) Cross-section of a sandstone showing small-scale
lamination in 'subcritically' (as defined in \protect\cite{hunter1}
and \protect\cite{allen}) inverse-graded ripples (along with the
quantities defined in the model).  Each pair of layers of small and
large grains is produced by the climbing of a single ripple.  ({\bf
b}) Example of cross-stratification.  Each climbing ripple produces a
series of layers of small and large grains oriented across the
direction of climbing, parallel to the downwind face.  The basic types
of the smallest stratification structures in ripples and small aeolian
dunes relevant to this study are summarized in
\protect\cite{hunter1}.}
\label{climb}
\end{figure}

Climbing of ripples as seen in Fig. \ref{diagram} is described by a
vector of translation which has two components \cite{hunter2}: the
component in the horizontal direction is the rate of ripple migration
across the sediment surface, and the component in the vertical
direction is the net rate of deposition, defined as the rate of
displacement of the sand surface in the vertical direction (see
Fig. \ref{climb}a). The angle of climbing ($\xi$ in Fig. \ref{climb}a)
defines different types of lamination structures (see Chapter 9
\cite{allen}).  If the angle of climb is smaller than the slope of the
stoss, then the structures are called ``subcritically'' climbing
ripples, and these are the sedimentary structures which are the focus
of this study.

Due to the climbing of ripples and segregation of grains during
 deposition and transport, ripple deposits develop lamination
 structures which are later preserved during solidification of the
 rock
\cite{hunter1,hunter2,allen}.  The basic types of the smallest
stratification structures in climbing ripples and small aeolian dunes
relevant to this study are summarized by Hunter \cite{hunter1} (see
also \cite{allen}).  In this paper we will focus on two lamination
structures which are commonly found in sedimentary rocks:

\begin {itemize}
\item
Inverse-graded climbing ripple lamination: one of the most common
lamination structures which are formed because grains composing the
ripples differ in size \cite{bagnold,hunter1,allen,bunas}.  Large
ejected grains travel in shorter trajectories than small grains, so
that small grains are preferentially deposited in the trough of the
ripples while large grains are deposited preferentially near the
crest.  Due to this segregation effect, the migration of a single
ripple produces two layers of different grain size parallel to the
climbing direction of the ripple as shown in Fig. \ref{climb}a. This
lamination structure is called inverse-graded--- according to the
nomenclature of Hunter \cite{hunter1}--- since in a pocket of two
consecutive layers formed by the climbing of a single ripple, the
layer of large grains is above the layer of small grains
(Fig. \ref{climb}a).

\item Cross-stratification: Migration of ripples also produces
successive layers of fine and coarse grains not in the direction of
climbing but parallel to the downwind face of the ripples (see
Figs. \ref{diagram} and \ref{climb}b)
\cite{bagnold,hunter1,reineck,hunter2,allen,kocurek}.  These
structures--- called cross-stratification or foresets--- are also
coarser toward the troughs \cite{bagnold,brown,williams,makse1} as
opposed to the segregation in inverse-graded climbing ripples.

\end{itemize}

The instability giving rise to aeolian ripple morphologies has been
the subject of much study.  The classic work of Bagnold \cite{bagnold}
has been followed up by a number of studies.  Modern models of wind
ripple deposits are usually defined in terms of the splash functions
introduced by Haff and coworkers \cite{splash,haff1} (see also
\cite{anderson2} for a review) in order to model impact processes.
These models have successfully reproduced the instability leading to
ripple deposits when the sand grains are of only one size
\cite{anderson2,haff2,anderson1}.  When such models are defined for
two species of grains differing in size \cite{bunas}, they result in
patterns which resemble very closely the stratigraphic patterns found
in inverse grading climbing ripple structures.  It was in this context
that Forrest and Haff proposed \cite{haff2} that grading changes in
ripple lamination are related to fluctuations in wind or transport.
However, Anderson and Bunas \cite{bunas} found using a cellular
automaton model that inverse-graded ripple lamination is due to
different hopping lengths of small and large grains.  These models do
not include the interactions between the moving grains and the static
surface \cite{anderson1}--- i.e., grains are assumed to stop as soon
as they reach the sand surface--- which are expected to be relevant
for the dynamics in the rolling face of the ripples
\cite{landry,terzidis}.  Recent studies by Terzidis {\it et al.}
\cite{terzidis} have reproduced the ripple instability using the
theory of surface flows of grains
proposed by Bouchaud and collaborators \cite{bouchaud}.  This theory
takes into account the interaction between rolling and static grains.
Moreover, recent studies of avalanche segregation of granular mixtures
show dramatic effects when such interactions are taken into account
\cite{makse2,bdg}.

\begin{figure}
\centerline{ \hbox{ \epsfxsize=7.cm \epsfbox{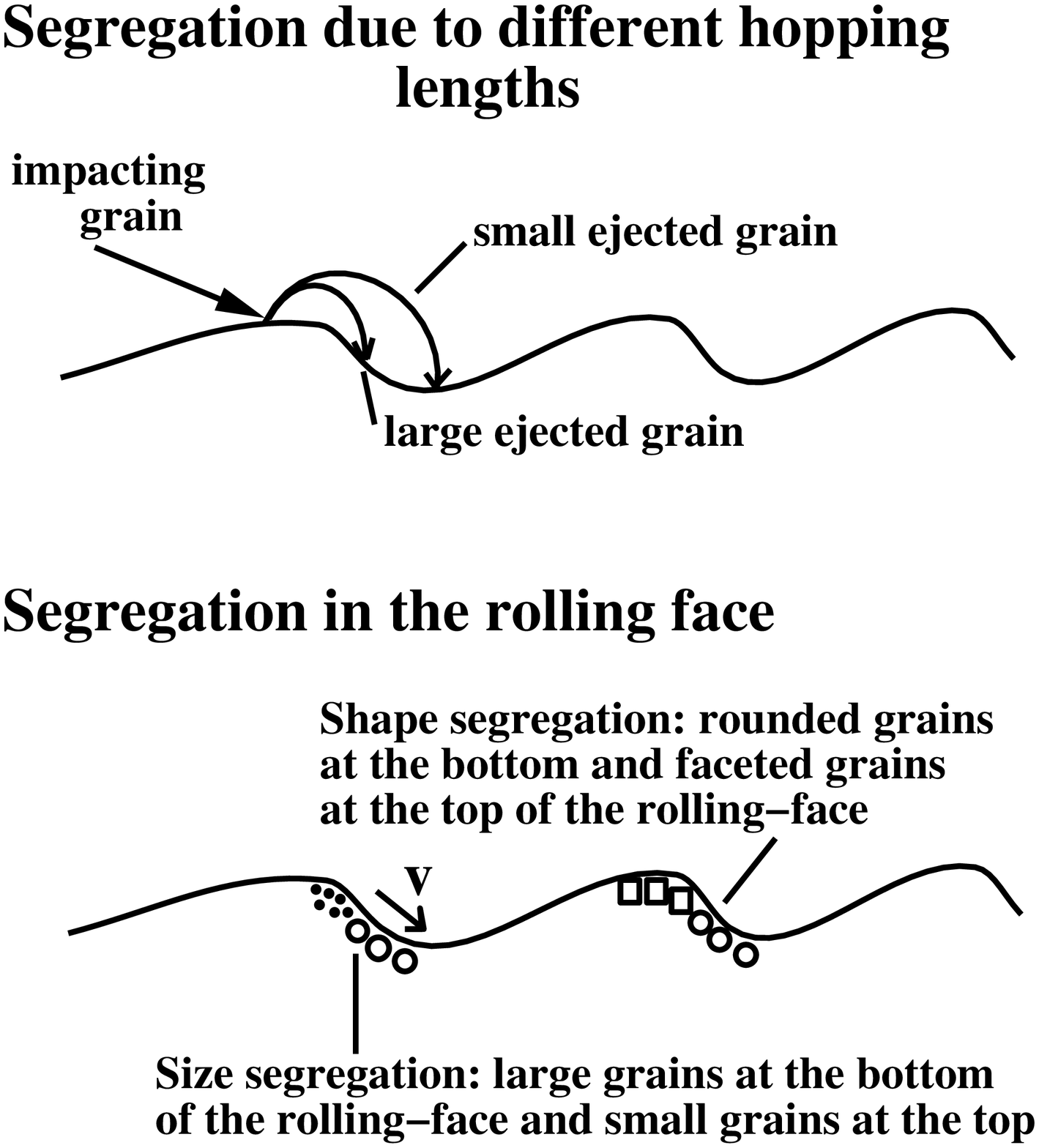} } }
\vspace{.5cm}
\caption{Three segregation mechanism acting when the grains differ in
size and shape.}
\label{mechanism}
\end{figure}

In this article, we use this formalism to include also the segregation
effects arising when considering two type of species of different size
and shape.  We first formulate a discrete model for two-dimensional
transverse aeolian climbing ripples which incorporates simple rules
for hopping and transport.  Then, we incorporate the different
properties of the grains, such as size and roughness, and we show that
pattern formation in ripple deposits arises as a consequence of
``grain segregation'' during the flow and collisions.  Specifically we
show that three segregation mechanisms contribute to layer formation
during ripple migration (Fig. \ref{mechanism}):

\begin{itemize}

\item Size segregation due to different hopping lengths of small and
large grains: the hopping length of the large reptating grains is
smaller than the hopping length of small grains.

\item Size segregation during transport and rolling: larger grains
tend to roll down to the bottom of troughs while small grains tend to
be stopped preferentially near the crest of ripples.

\item Shape segregation during transport and rolling: rounded grains
tend to roll down easier than more faceted or cubic grains, so that
the more faceted grains tend to be at the crest of the ripples, and
more spherical grains tend to be near the bottom.

\end{itemize}

As a result of a competition between these segregation mechanisms a
richer variety of stratigraphic patterns emerges: inverse-graded
ripple lamination occurs when segregation due to different jump
lengths dominates, and cross-stratification (and normal-graded ripple
lamination as well) occurs when size and shape segregation during
rolling dominates.  Thus, a general framework is proposed to unify the
mechanisms underlying the origin of the most common wind ripple
deposits.  In the case of inverse-graded climbing ripple lamination we
confirm a previously formulated hypothesis by Bagnold \cite{bagnold}
and Anderson and Bunas \cite{bunas} that the lamination is due to
hopping induced segregation due to collisions.  Regarding the origin
of cross-stratification structures, we show that they arise under
similar conditions as in recent table-top experiments of avalanche
segregation of mixtures of large faceted grains and small rounded
grains poured in a vertical Hele-Shaw cell \cite{makse1,makse2}.

In what follows we first define the model for the case of single
species of grains. We show that the model predicts the formation of a
ripple structure, and propose a simplified continuum theory to
understand the onset of the stability leading to ripple structure.
Then, we generalize to the case of two type of species of grains
differing in size and shape which give rise to the segregation
mechanisms and the lamination structures seen in aeolian climbing
ripples.

\section{One species model}

We start by defining the model for the case of one type of grain in a
two-dimensional lattice of lateral size $L$ with periodic boundary
conditions in the horizontal $x$-direction.  Our main assumption is to
consider two different phases \cite{terzidis,bouchaud,makse2,bdg,pgg}:
a reptating or rolling phase composed by grains moving with velocity
$v$ by rolling, and a static phase composed by grains in the bulk.  We
also consider a curtain of external saltating grains which impact---
at randomly chosen positions on the static surface--- from the left to
the right at a small angle $\beta$ to the horizontal
(Fig. \ref{climb}a).  Shadow effects are considered by allowing only
impacts with ballistic trajectories which do not intersect any prior
portion of the surface.  Upon impacting on the sand surface, a
saltating grain dislodges $n_{rep}$ grains from the static surface,
and jumps a distance $l_{sal}$ after which the saltating grain is
incorporated to the reptating phase.  The $n_{rep}$ grains dislodged
by the saltating grain jump a distance $l_{rep}\ll l_{sal}$.  Upon
reaching the surface, the dislodged grains form part of the reptating
phase and move with velocity $v$.

At every time step ($\Delta t$) only one reptating grain (in contact
with the surface) interacts with the static grains of the bulk
according to the angle of repose $\theta_r$ \cite{coarse-grain}; the
remaining reptating grains move a distance $v \Delta t$ to the right.
The angle of repose is the maximum angle at which a reptating grain is
captured on the sand bed.  If the local coarse-grained angle of the
surface is smaller than the angle of repose, $\theta < \theta_r$, the
interacting reptating grain will stop and will be converted into a
static grains.  If the angle of the surface is larger than the angle
of repose, $\theta > \theta_r$, the reptating grain is not captured---
and moves to the right a distance $v \Delta t$ with the remaining
reptating grains--- but ejects a static grain from the bulk into the
reptating phase.

\begin{figure}
\centerline{ \hbox{ \epsfxsize=8cm \epsfbox{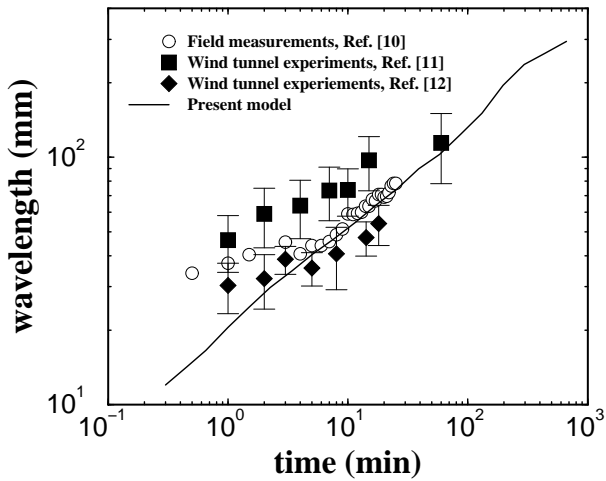} } }
\caption{Growth of the wavelength of the ripples according to our
model and comparison with field and experimental observations. The
number of impacts in the model is translated to time using a typical
impact rate of $10^7$ impacts $m^{-2}$ $s^{-1}$
\protect\cite{anderson2}.  The simulation data have been shifted
vertically by a multiplicative factor, so that only the general trend
(the slope) of the curve should be compared with the experimental
data.}
\label{wavelength}
\end{figure}

The model predicts the formation of a ripple structure as seen in
aeolian sand formations.  The onset of the instability leading to
ripples occurs with ripples of small wavelength.  However, due to the
fact that smaller ripples travel faster than larger ripples--- since
smaller ripples have less amount of material to transport--- merging
of small ripples on top of large ripples is observed.  This gives rise
to the change of the characteristic wavelength as a function of time,
as is observed in wind tunnel experiments \cite{seppala,anderson2}, as
well as in field observations \cite{sharp,werner}.  We find that our
model predicts an initial power law growth of the wavelength of the
ripples $\lambda(t) \sim t^{0.4}$ (see Fig. \ref{wavelength}).  The
wavelength of the ripples seems to saturate after this power law
growth to a value determined by the saltating length.  Moreover, we
also observe a subsequent growth of the wavelength (not shown in the
figure) up to a new saturation value close to a multiple of the
saltation length.  This process continues and a series of plateaus are
observed until one large ripple (or dunes) of the size of the system
is formed.  The series of plateaus was also observed in a recent
lattice model of ripple merging \cite{vandewalle}. However, we believe
that the existence of plateaus might be due to an artifact of the
discreteness of the lattice, and may have no physical meaning: the
rules associated with the derivatives of the sand surface break down
when the local slope is bigger than a given value determined by the
discretization used to define the slope of the surface. A similar
artifact was observed in a discrete model of ripple morphologies
formed during surface erosion via ion-sputtering in amorphous
materials \cite{rodolfo}.  Moreover, we will see that when we
introduce two species of grains, the wavelength saturates at a
constant value and the plateaus disappear.

In Fig. \ref{wavelength} we compare the prediction of our model with
the available experimental data from field and wind tunnel
experiments.  We observe a fair agreement between our model and the
experiments.  However, we believe that this agreement is not
conclusive.  In fact, the same experimental data can be also fitted
with the same accuracy by a logarithmic growth of the wavelength of
the ripples as shown recently by Werner and Gillespie \cite{werner},
who proposed a discrete stochastic model of ripple merging and found a
logarithmic growth of the wavelength (see also \cite{vandewalle}).

\subsection{Continuum formulation of the model}

The onset of the instability leading to the initial ripple structure
can be studied using the continuum theory proposed by Bouchaud {\it et
al.}  \cite{bouchaud} to study avalanche in sandpiles.  Recently, the
set of coupled equations for surface flow of grains of \cite{bouchaud}
has been adapted to the problem of the ripple instability
\cite{terzidis}. Here, we propose a version of this theoretical
formalism suitable to the physics of our model to study the initial
ripple formation.  Let $R(x,t)$ describe the amount of reptating
grains in the rolling phase and $h(x,t)$ the height of the static bed
at position $x$ and time $t$, which is related to the angle of the
sand surface by $\theta(x,t) \equiv -\partial h/\partial x$ for small
angles.  Our set of convective-diffusion equations for the rolling and
static phases are the following \cite{bouchaud}:

\begin{mathletters}
\begin{equation}
\frac{\partial R(x,t) }{\partial t} = - v \frac{\partial R}{\partial
x} + D \frac{\partial^2 R}{\partial x^2} + \Gamma(R,\theta),
\label{r}
\end{equation}

\begin{equation}
\frac{\partial h(x,t)}{\partial t} = - \Gamma(R,\theta),
\label{h}
\end{equation} 
\label{rh}
\end{mathletters}
\noindent
where $v$ is the drift velocity of the reptating grains along the
positive $x$ axis taken to be constant in space and time, and $D$ a
diffusion constant.  The interaction term $\Gamma$ takes into account
the conversion of static grains into rolling grains, and vice
versa. We propose the following form of $\Gamma$ consistent with our
model:
 
\begin{equation}
\Gamma(R,\theta) \equiv \alpha [\beta - \theta(x,t) ] + \gamma [
\theta(x,t) - \theta_r ] R(x,t),
\label{gamma}
\end{equation}
where $\alpha$ and $\gamma$ are two constants with dimension of
velocity and frequency respectively. $\alpha$ is proportional to the
number of collisions per unit time of the saltating grains with the
sand bed, and also to the number of ejected grains per unit time,
while $\gamma$ is the number of collisions per unit time between the
reptating grains and the sand bed when they creep on the downwind side
of the ripple.  The first term in (\ref{gamma}) takes into account the
spontaneous (independent of $R(x,t)$) creation of reptating grains due
to collisions of the saltating grains--- a process which is most
favorable at angles of the bed smaller than $\beta$.  The second term
takes into account the interaction of the reptating grains with the
bed of static grains: the rate of the interaction is proportional to
the number of grains $R(x,t)$ interacting with the sand surface.
Rolling grains become part of the sand surface if the angle of the
surface $\theta(x,t)$ is smaller than the repose angle $\theta_r$
(``capture''), while static grains become rolling grains if
$\theta(x,t)$ is larger than $\theta_r$ (``amplification'').  Higher
order terms are neglected since we are interested in the linear
stability analysis determining the origin of the ripple instability.

Insight into the mechanism for ripple formation is obtained by
studying the stability of the uniform solution of Eqs. (\ref{rh}):
\begin{equation}
R_0=\alpha\beta / (\gamma \theta_r) ~~~~~~~ h_0=0.
\end{equation} 

We perform a stability analysis by looking for solutions of the type:
$R(x,t) = R_0 + \bar{R}(x,t)$ and $\theta(x,t) = \theta_0 + \bar
\theta(x,t)$.  The linearized equations for $\bar R$ and $\bar \theta$
are

\begin{mathletters}
\begin{equation}
\displaystyle{ \frac{\partial \bar R(x,t) }{\partial t}
}=\displaystyle{ - v \frac{\partial \bar R} {\partial x} - v_m \bar
\theta - \gamma \theta_r \bar R + D \frac{\partial^2 \bar R}{\partial
x^2},}
\end{equation}
\begin{equation}
\displaystyle{ \frac{\partial \bar \theta(x,t) }{\partial t}}
=\displaystyle{ - v_m ~\frac{\partial \bar \theta} {\partial x} -
\gamma \theta_r \frac{ \partial \bar R}{\partial x}, }
\end{equation}
\label{bar}
\end{mathletters}
where the migration velocity of the traveling wave solution is
\begin{equation}
v_m = \alpha (1-\beta/\theta_r),
\label{trans}
\end{equation}
which indicates that the ripple velocity is proportional to the number
of collisions per unit unit of saltating grains with the sand bed, and
to the number of ejected grains per collision.

By Fourier analyzing Eqs. (\ref{bar}) we find a set of two homogeneous
algebraic equations for $\bar R$ and $\bar\theta$ with non-trivial
solutions only when the determinant of the resulting $2 \times 2$
matrix is zero.  We obtain a dispersion relation $w_\pm(k)$ where the
two branches $\pm$ correspond to the solutions of the resulting
quadratic equation for $w(k)$.  Then we take the limit $v_m/v \ll 1$,
which corresponds to the physical fact that the translation velocity
(\ref{trans}) is much smaller than the rolling velocity of the
individual grains (see \cite{terzidis}), and we arrive to the
following dispersion relation for $w_{\_}(k)$ ($w_+(k)$ gives rise
only to stable modes):

\begin{equation}
\mbox{Im} [w_{\_}(k)] = \frac {-(v_m/v)~ (\gamma \theta_r v^2)~k^2 }
 {(\gamma\theta_r)^2 + k^2 (2 \gamma\theta_r D + v^2) + D^2 k^4} +O(
 \frac{v_m}{v})^2.
\end{equation}

The asymptotic forms for small and large $k$ are
\begin{equation}
\begin{array}{ll}
\displaystyle{ \mbox{Im} [w_{\_}(k)] \rightarrow - \frac {v_m
v}{\gamma \theta_r} k^2, } & \mbox{~~~$k\to 0$}\\ \displaystyle{
\mbox{Im} [w_{\_}(k)] \rightarrow - \frac {v_m v \gamma \theta_r}{D^2}
\frac{1}{k^2}, } & \mbox{~~~$k\to \infty$}
\end{array}
\end{equation}
which indicates that the branch $w_{\_}(k)$ corresponds to only
unstable modes $w_{\_}(k) < 0$. Similar stability analysis was
performed by Anderson \cite{anderson1} using a continuity model
neglecting the interaction between rolling and static grains and by
Terzidis {\it et al.}  \cite{terzidis} with the continuum model
Eqs. (\ref{rh}) but with a different interaction term than
(\ref{gamma}). Terzidis {\it et al.} find a band of unstable modes
until a given cut off wave vector at large $k$.  This behaviour is due
to higher order derivatives of $\theta$ appearing in the interaction
term used in \cite{terzidis}.  The most unstable mode $k^*$ in our
model is given by $\partial w_{\_}(k^*)/\partial k = 0$, with $k^* =
\sqrt{\gamma \theta_r/D}$ which gives an estimate of the initial
wavelength of the ripples:
\begin{equation}
\lambda = 2\pi \sqrt{D/(\gamma \theta_r)}.
\end{equation}

The final wavelength will be determined by higher order nonlinear
terms arising from a more complicated interaction term than the one
used in (\ref{gamma}). As mentioned above, using the full discrete
model we find that after the appearance of initial small undulations,
the wavelength of the ripples grows due to ripple merging. In the
discrete model, the wavelength seems to saturate to a finite value,
although this value seems to be determined by the finite size of the
simulation system (we use periodic boundary conditions in the
horizontal direction).


\section{A model for two species differing in size and shape}

Next we generalize the model to the case of two type of species
differing in size and shape. The segregation mechanisms discussed in
the introduction are incorporated in the model as follows:

\begin {itemize}
\item
Size segregation due to different hopping lengths: we define
$l_{\alpha}$ (to replace $l_{rep}$ of the mono disperse case) as the
distance a reptating grain of type $\alpha$ travels after being
collided by a saltating grain.  If we call the small grains type $s$
and the large grains type $l$, then $l_{l}<l_{s}$.  This effect was
incorporated in the discrete stochastic model of Anderson and Bunas
\cite{bunas} using a generalization of the splash functions proposed
by Haff \cite{haff1}.
\end{itemize}

\noindent The interaction between the rolling grains and the surface
is characterized by four different angles of repose;
$\theta_{\alpha\beta}$ for the interaction of an $\alpha$ reptating
grain and a $\beta$ static grain (replacing $\theta_r$ of the mono
disperse case).  The dynamics introduced by the angle of repose are
relevant at the downwind face where two extra mechanisms for
segregation of grains act in the system:

\begin{itemize}

\item Size segregation during transport and rolling: large rolling
grains are found to rise to the top of the reptating phase while the
small grains sink downward through the gaps left by the motion of
larger grains; an effect known as percolation or kinematic sieving
\cite{bagnold,drahun-savage}.  Due to this effect only the small
grains interact with the surface when large grains are also present in
the rolling phase. Small grains are captured first near the crest of
the ripples causing the larger grains to be convected further to the
bottom of the ripples.  We incorporate this dynamical segregation
effect by considering that, when large and small grains are present in
the reptating phase, only the small ones interact with the surface
according to the angle of repose, while the large grains, being at the
top of the reptating phase do not interact with the static grains and
they are convected downward.
Thus, the large grains interact with the surface only when there are
no small grains present in the reptating phase.  A similar percolation
mechanism was introduced in the discrete and continuum models of
\cite{makse2} to understand the origin of stratification patterns in
two-dimensional sandpiles of granular mixtures.

\item Shape segregation during transport and rolling: rounded grains
tend to roll down easier than more faceted or cubic grains, so that
the more faceted grains tend to be at the crest of the ripples.  This
segregation effect is quantified by the angles of repose of the pure
species, since the repose angle is determined by the shape and surface
properties of the grains and not by their size: the rougher or the
more faceted the surface of the grains the larger the angle of repose.
If the large grains are more faceted than the small grains we have
$\theta_{ss}< \theta_{ll}$, while when the small grains are more
faceted $\theta_{ll} < \theta_{ss}$ \cite{makse2}.  The species with
the larger angle of repose have a larger probability to be captured at
the sand bed than the species with smaller angle of repose.

\end{itemize}

We notice that the fact that the grains have different size leads to
different cross-angles of repose $\theta_{\alpha\beta}$
\cite{makse2,bmdg}.  If the large grain are type $l$ and the small are
type $s$ then $\theta_{ls}< \theta_{sl}$, which in turn leads to
another size segregation effect.  However, by incorporating the size
segregation due to the different cross-angle of repose we would get
the same effect as the one incorporated in the model by the
percolation effect (the small grains are preferentially trapped at the
top of the slip-face), so that, in what follows, we will consider only
the percolation effect for simplicity--- i.e., large reptating grains
are allowed to interact with the sand surface only when there are no
small reptating grains below the large grains at a given position.

We also notice that, in general, the distance a grain travels after
being kicked by a saltating grain depends on the type of colliding
grain and the type of grain on the bed.  Thus we define
$l_{\alpha\beta}$ as the distance a reptating grain of type $\beta$
travel after being collided by a saltating grain of type $\alpha$. If
we call the large grains type $l$ and the small grains type $s$ then
we have: $l_{sl}<l_{ll}<l_{ss}<l_{ls}$.  Hoverer, in practice we will
make the following approximation $l_{sl}=l_{ll}\equiv l_l$, and
$l_{ss}=l_{ls}\equiv l_s$, i.e., we do not consider which type of
grain is colliding.

Thus, three different mechanisms--- size-segregation due to
percolation in the reptating phase, shape-segregation in avalanches,
and size segregation due to the different reptating jumps
$l_{\alpha}$--- compete in the system giving rise to a rich variety of
lamination patterns as we show below.

\subsection{Size segregation due to different hopping lengths}

\begin{figure}
\centerline{ \hbox{ \epsfxsize=8.5cm \epsfbox{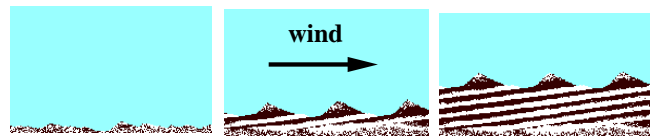} } }
\vspace{.5cm}
\caption{Morphology predicted by the model when segregation due to
different jump lengths is dominant showing inverse grading climbing
ripple lamination.  From left to right, we show a sequence of three
stages in the dynamics of the climbing ripples.  Starting from a flat
sand surface of small (light color) and large grains (dark color), the
system evolves into ripples climbing in the direction of the wind.}
\label{seq}
\end{figure}

We first investigate the morphologies predicted by the model in the
case where the different reptating jumps $l_{\alpha}$ play the
dominant role in the segregation process.  This is the case
$l_{s}-l_{l}>l_s$ and when the grains have approximately the same
shape $\theta_{ss}\approx\theta_{ll}$.  We start our simulations from
a flat sand surface composed by a 50/50 by volume mixture of small and
large grains and we observe the system to evolve into ripples
traveling in the direction of the wind (Fig. \ref{seq}).  The
resulting morphology seen in Fig. \ref{seq} resembles the most common
climbing ripple structures such as those documented in field
observations \cite{hunter1,hunter2,allen,kocurek}
(Fig. \ref{climb}a). This result confirms the hypothesis of Anderson
and Bunas \cite{bunas} that the origin of inverse-graded lamination in
climbing ripples is due to the size segregation effect produced by the
different hopping lengths of small and large grains.  In our
simulations we observe that large grains (dark color) are deposited at
the top of the ripples while small grains (light color) are deposited
preferentially at the bottom of the ripples since $l_{l}<l_{s}$,
resulting in a lamination structure showing inverse grading (layer of
large grains on top of the layer of small grains).

The system size used in our simulations corresponds to 256 bins, with
$\Delta t=1$ s, $n_{rep}=4$ grains per impact, $l_{sal}=90$ cm, $\beta
= 10^o$.  For the case shown in Fig. \protect\ref{seq} we use $l_1 =
25$ cm and $l_2= 5$ cm, and $\theta_{\alpha\beta}= 30^o$.

The efficiency of the segregation mechanisms is sensitive to the
relative value of the parameters of the model.  For instance we find
that by decreasing the value of $l_{sal}$ relative to the reptating
lengths, the efficiency of segregation is greatly reduced in the case
of inverse-graded lamination.  The structures shown here are all
subcritically climbing from left to right (for a review of the
terminology and definition of different climbing structures in ripples
deposit see \cite{hunter1}, and Chapter 9 \cite{allen}).
Supercritical climbing ripples are much less common and are produced
in slowly translating ripples with small number of ejected grains per
impact \cite{haff2}.  We also notice that the shape of ripples found
in Nature are more asymmetric with the downwind side at a steeper
angle than the upwind side, while our model predicts a more symmetric
triangular shape of the ripples.  More realistic asymmetric shape of
the ripples can be obtained by considering the exact trajectories of
the flying grains and the complex interaction of grains with the air
flow as shown by previous models by Anderson \cite{bunas}.

\subsection{Segregation due to rolling and transport}

Next we consider the case where the difference between the reptating
jumps is small, $|l_{s}-l_{l}|/l_{s}< 1$ (or when the downwind face is
large compared to $l_{s}$ or $l_{l}$) and the grains differ
appreciable in shape ($\theta_{ss}\ne \theta_{ll}$) and also in size.
Then segregation in the rolling face is the relevant mechanism for
segregation and we do not take into account the segregation due to
different hoping lengths.

\begin{figure}
\centerline{ \hbox{ \epsfxsize=9cm \epsfbox{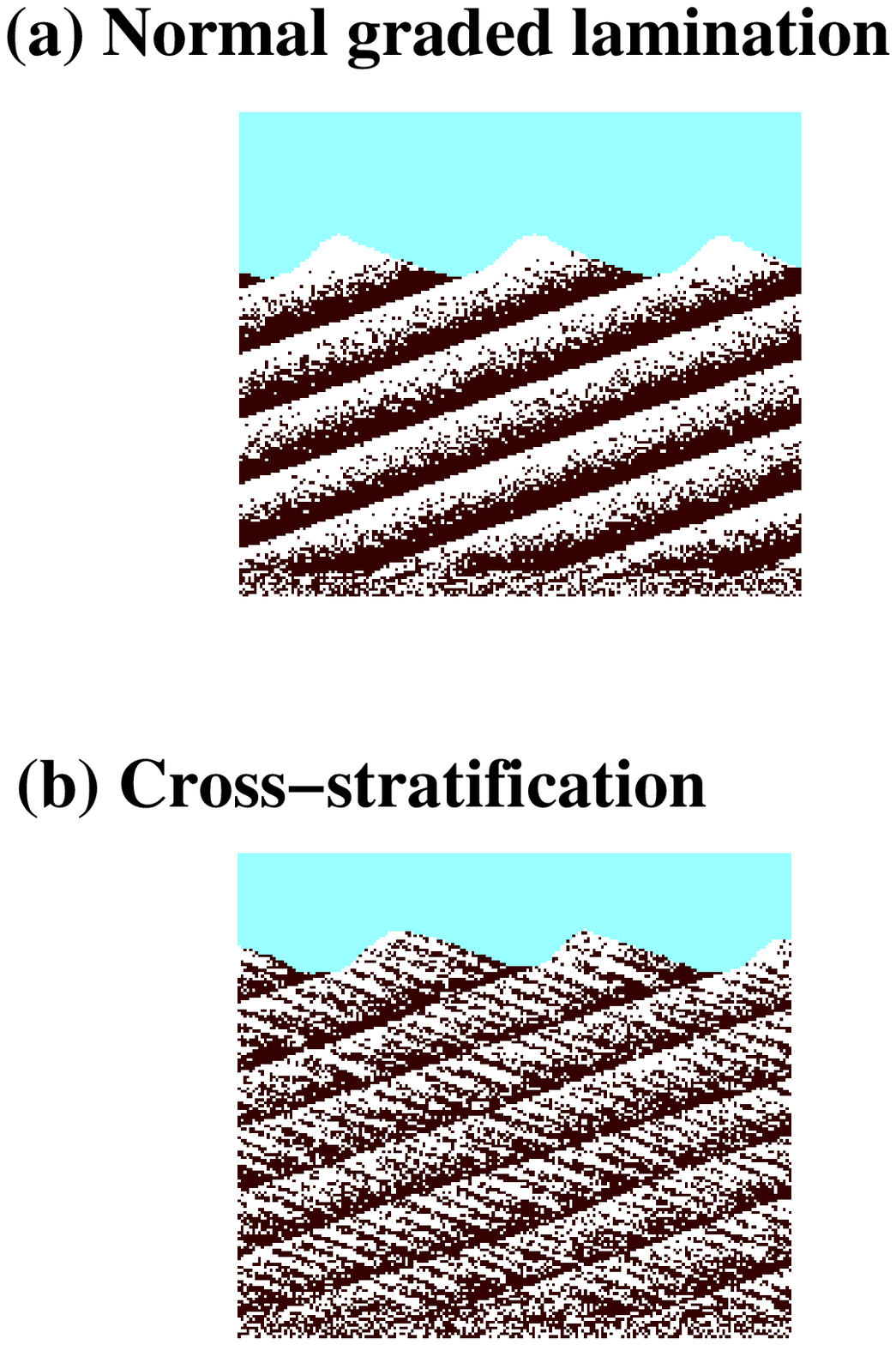} } }
\vspace{.5cm}
\caption{Resulting morphologies predicted by the model after $10^7$
impacts, when segregation in the rolling face is dominant showing
({\bf a}) normal grading lamination of large rounded grains (dark
color, at the bottom of the downwind face) and small rough grains
(light color, at the top); and ({\bf b}) cross-stratification of large
rough grains (dark color) and small rounded grains (light color).}
\label{phase}
\end{figure}

We first consider the case where the small grains are the roughest and
the large grains are the most rounded ($\theta_{ss} > \theta_{ll}$).
In this case a segregation solution along the downwind face is
possible since both, size segregation due to percolation and shape
segregation act to segregate the small-faceted grains at the top of
the crest and the large-rounded grains at the bottom of troughs.  The
result is a lamination structure (Fig.  \ref{phase}a) which resembles
the structure of climbing ripples of Fig. \ref{seq} but with the
opposite grading: small grains at the top of each pair of layers
(normal-grading climbing ripple lamination). This type of lamination
is not very common in Nature.

On the other hand, when the large grains are rougher than the small
grains ($\theta_{ll} > \theta_{ss}$), a competition between size and
shape segregation occurs \cite{makse3}.  Size segregation due to
percolation tends to segregate the large-cubic grains at the bottom of
the downwind face, while the shape segregation effect tends to
segregate the same grains at the crest of the ripples. Then a
segregation solution along the downwind face as in Fig. \ref{phase}a
is not possible, and the result of this instability is the appearance
of layers of small and large grains parallel to the downwind face
(Fig. \ref{phase}b) and not perpendicular as in Fig.  \ref{phase}a.
These structures correspond to the more common cross-stratification
patterns found in rocks \cite{hunter1,hunter2,allen,kocurek}.  In
addition to the stratification parallel to the rolling face, the
deposits are also coarser toward the bottom of the downwind face.

The mechanism for cross-stratification involves two phenomena
superimposed.  {\it (i)} Segregation in the rolling face: a pair of
layers is laid down in a single rolling event, with the small grains
segregating themselves underneath the larger grains through a ``kink''
mechanism as seen in \cite{makse1,makse2}.  This kink in the local
profile of static grains provides local stability to the rolling
grains trapping them. The kink moves uphill forming layers of small
and large grains.  {\it (ii)} The climbing of ripples due to grain
deposition and ripple migration.

The conditions for cross-stratification and the dynamical segregation
process found with the present model are similar to the findings of
the experiments of stratification of granular mixtures in
two-dimensional vertical cells performed in \cite{makse1} and also the
models of \cite{makse2}.  We observe the formation on the slip face of
the ripples of an upward traveling wave or ``kink'' at which grains
are stopped during the avalanche as was observed in
\cite{makse1,makse2}.  According to this, the wavelength of the layers
is proportional to the flux of grains reaching the rolling face, then
it is proportional to the number of saltating grains impacting the
surface.

For the case shown in Fig. \protect\ref{phase}a we use $l_s = l_l = 5$
cm, and $\theta_{sl}=\theta_{ss} = 30^o$, and $\theta_{ls}=\theta_{ll}
= 20^o$, while for Fig. \protect\ref{phase}b we use $l_s = l_l = 5$
cm, and $\theta_{sl}=\theta_{ss} = 20^o$, and $\theta_{ls}=\theta_{ll}
= 30^o$.


\subsection{Phase diagram: General case}

\begin{figure}
\centerline{ \hbox{ \epsfxsize=9cm \epsfbox{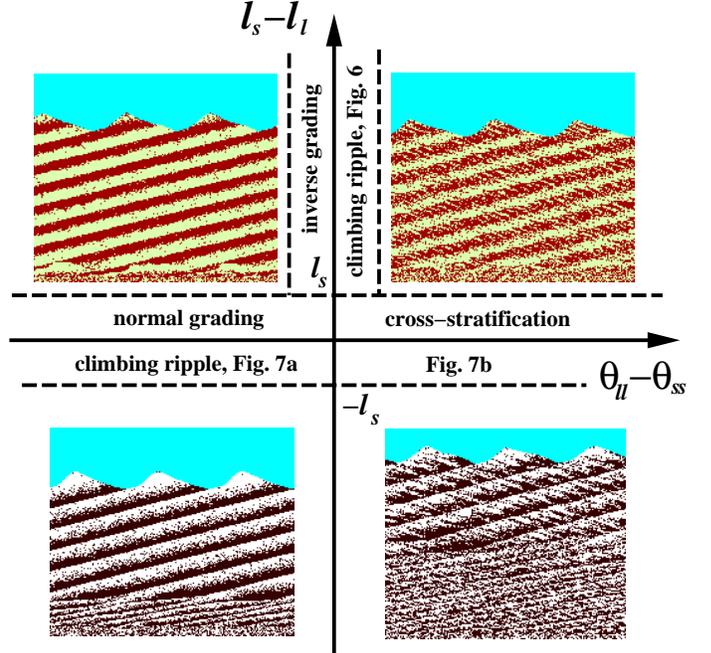} } }
\caption{Phase diagram predicted by the model.  The ``$s$'' refers to
the small grains (light colors), and the ``$l$'' refers to the large
grains (dark colors).  $\theta_{ss}>\theta_{ll}$ means that the
smaller grains are the roughest, and $\theta_{ll}>\theta_{ss}$ means
that the larger grains the roughest.  The three areas in the phase
space near the axis refer to the morphologies already studied in
Figs. \protect\ref{seq}, \protect\ref{phase}a, and
\protect\ref{phase}b.}
\label{phase2}
\end{figure}

We have also investigated the morphologies obtained when the three
segregation effects (shape segregation, percolation effect, and
hopping induced size segregation) act simultaneously.  The resulting
morphologies are shown in Fig. \ref{phase2} along with the phase
diagram summarizing the results obtained with the model in Figs.
\ref{seq} and \ref{phase}.

The case $|l_{s}-l_{l}|>l_s$ and $\theta_{ss}\approx\theta_{ll}$
corresponds to the inverse-graded lamination shown in Fig. \ref{seq},
as indicated in the phase diagram Fig. \ref{phase2}.  Moreover, when
$l_{s}-l_{l}>l_s$ and for any other value of the angles $\theta_{ss}$
and $\theta_{ll}$, we find that the hopping induced segregation seems
to dominate over rolling induced segregation.  As can be seen from the
upper two panels in Fig.  \ref{phase2}, the morphologies obtained in
these cases resemble the one of Fig.  \ref{seq} corresponding to
inverse-graded climbing ripple lamination.  The upper left panel shows
the results when the smaller grains are the roughest
$\theta_{ll}-\theta_{ss}<0$ and shows clearly the same lamination as
in Fig. \ref{seq}. In this case the hopping induced segregation
dominates completely over the rolling segregation due to the angles of
repose and percolation.  In the upper right panel we show the case
when the large grains are the roughest $\theta_{ll}-\theta_{ss}>0$. In
this case we also see the same lamination pattern as in Fig. \ref{seq}
but we see a tenuous trace of the cross-stratification seen in
Fig. \ref{phase}b too; the rolling induced segregation seems to have a
more important role in comparison with the hopping induced segregation
than in the case shown in the upper left panel.

The region $|l_{s}-l_{l}|<l_{s}$ is discussed in Figs. \ref{phase},
and our model shows normal grading lamination when
$\theta_{ll}-\theta_{ss}<0$ (Fig. \ref{phase}a) and
cross-stratification when $\theta_{ll}-\theta_{ss}>0$
(Fig. \ref{phase}b).

Interesting morphologies are predicted when $l_{s}-l_{l}<-l_s$.  This
might be the case of a large difference in density between the grains,
i.e., very light large grains and heavy small grains. The patterns
obtained with the model are shown in the two lower panels in Fig.
\ref{phase2}.  The late stage in the dynamics evolution of the
patterns shown at the left lower panel ($\theta_{ll}-\theta_{ss}<0$)
and at the right lower panel ($\theta_{ll}-\theta_{ss}>0$) in
Fig. \ref{phase2} resemble the patterns of normal grading climbing
ripples (Fig. \ref{phase}a) and cross-stratification
(Fig. \ref{phase}b) respectively. This indicates the dominance of
avalanche segregation (shape segregation and percolation effect) over
the hopping induced segregation in this region of the phase space
(this dominance is the opposite of what we find in the region
$l_{s}-l_{l}>l_{s}$).  However, as seen in Fig. \ref{phase2}, these
patterns appear only in the late stages of the evolution.  As seen in
the two lower panels, in the early stages the lamination structure
appears to be dominated by the hopping induced segregation.  We
observe a clear transition at intermediate time in the simulation
which can be clearly observed in the change of the climbing angle of
the ripples observed in the left and right lower panels in
Fig. \ref{phase2}.  The late stage patterns appear only when the
slip-face is well developed after a transient of small ripples
dominated by hopping segregation.  We are not aware of any wind tunnel
experiment or field observation showing similar dynamical transition,
so it would be interesting to explore this transition experimentally.

\section{Summary}

In summary, we have shown that lamination in sedimentary rocks at
around the centimeter scale is a manifestation of grain size and grain
shape segregation.  Thin layers of coarse and fine sand are present in
these rocks, and understanding how layers in sandstone are created
might aid, for instance, in oil exploration since great amounts of oil
are locked beneath layered rocks.  Our findings suggest a unifying
framework towards the understanding of the origin of inverse grading,
normal grading and cross-stratification patterns in wind ripple
deposits.  We identify the conditions under which these different
laminations patterns arise in sandstone. In particular, we find that
cross-stratification is only possible when the large grains are
coarser than the small grains composing the layered structure in
agreement with recent experiments on avalanche segregation in
two-dimensional sandpiles.  The fact that cross-stratification
patterns are very common in Nature might be due to the fact that
fragmentation processes usually leads to smaller grains more rounded
than larger grains.  So far we have explored a two-dimensional
cross-section of the ripple deposits.  Extensions to three dimensional
systems are ongoing as new physics may emerge when taking into account
the lateral motion of the grains \cite{landry}.


\end{multicols}


\begin{references}
\bibitem{coarse-grain} Notice that all the quantities of the model are
averaged over a coarse-grained scale of the order of a few grain
sizes.
\end{references}
\end{document}